\documentclass{article}
\usepackage{spconf,amsmath,graphicx}

\usepackage{color}
\usepackage{enumitem}
\usepackage{multirow}
\usepackage{etoolbox,siunitx}
\robustify\bfseries
\usepackage{booktabs}
\usepackage{balance}
\usepackage{amssymb}
\usepackage{bm}
\usepackage{arydshln}
\usepackage{caption}
\usepackage{cite}
\usepackage{hyperref}

\usepackage{amssymb}%
\usepackage{pifont}%
\newcommand{\cmark}{\ding{51}}%
\newcommand{\xmark}{\ding{55}}%

\def\T{{\mathsf T}}

\def\RR{{\mathbb R}}

\usepackage{caption}
\let\OLDthebibliography\thebibliography
\renewcommand\thebibliography[1]{
  \OLDthebibliography{#1}
  \setlength{\parskip}{0.5pt}
  \setlength{\itemsep}{1pt plus 0.3ex}
}
\title{TF-GridNet: Making Time-Frequency Domain Models Great Again\\for Monaural Speaker Separation}
\name{Zhong-Qiu Wang$^1$, Samuele Cornell$^{1,2}$, Shukjae Choi$^3$, Younglo Lee$^3$, Byeong-Yeol Kim$^3$, Shinji Watanabe$^1$}
\address{
$^1$Language Technologies Institute, Carnegie Mellon University, Pittsburgh, USA \\
$^2$Università Politecnica delle Marche, %
Italy \,\,\,\, $^3$Hyundai Motor Group and 42dot Inc., Seoul, Korea \\
{\small\texttt{wang.zhongqiu41@gmail.com}}\vspace{-0.4cm}
}
\begin{document}
\ninept

\maketitle
\setlength{\abovedisplayskip}{2pt}
\setlength{\belowdisplayskip}{2pt}
\begin{abstract}
We propose TF-GridNet, a novel multi-path deep neural network (DNN) operating in the time-frequency (T-F) domain, for monaural talker-independent speaker separation in anechoic conditions.
The model stacks several multi-path blocks, each consisting of an intra-frame spectral module, a sub-band temporal module, and a full-band self-attention module, to leverage local and global spectro-temporal information for separation.
The model is trained to perform complex spectral mapping, where the real and imaginary (RI) components of the input mixture are stacked as input features to predict target RI components.
Besides using the scale-invariant signal-to-distortion ratio (SI-SDR) loss for model training, we include a novel loss term to encourage separated sources to add up to the input mixture.
Without using dynamic mixing, we obtain 23.4 dB SI-SDR improvement (SI-SDRi) on the WSJ0-2mix dataset, outperforming the previous best by a large margin. %
\end{abstract}
\vspace{-0.1cm}
\section{Introduction}
\vspace{-0.1cm}

Dramatic progress has been made in monaural talker-independent speaker separation since the invention of deep clustering~\cite{Hershey2016} and permutation invariant training (PIT)~\cite{Kolbak2017}.
Early studies train DNNs to perform separation in the magnitude domain, with or without using magnitude-based phase reconstruction~\cite{Isik2016, Wang2018AlternativeObejectives, WZQe2eMISI2018, Wang2018i, Wang2019Trigonometric}.
Subsequent studies perform separation in the complex T-F domain through complex ratio masking~\cite{Liu2019DeepCASA} or in the time domain through the encoder-separator-decoder scheme proposed in TasNet~\cite{Luo2017TasNet, Luo2018TasNetRealTime, Luo2019}.
Since 2019, TasNet and its variants~\cite{Luo2019, Lam2020MBT, Shi2019FurcaNeXt, Tzinis2020, Luo2020, Nachmani2020, Chen2020DPTnet, Zhu2021, Subakan2021, Lam2021, Zeghidour2020, Qian2022, Rixen2022}, which feature learned encoder and decoder operating on very short windows of signals, have gradually become the most popular and dominant approach for speaker separation in anechoic conditions, largely due to their strong performance and advanced DNN architectures designed for end-to-end optimization.
Their performance on WSJ0-2mix~\cite{Hershey2016}, the de-facto benchmark dataset for speaker separation in anechoic conditions, has reached an impressive 22.1 dB SI-SDRi in a recent study~\cite{Rixen2022QDPN}.

In the meantime, T-F domain models, which typically use larger window sizes and hop sizes, have been largely under-explored and under-represented in anechoic speaker separation.
Recently, TFPSNet~\cite{Yang2022TFPSNet}, which claims to operate in the complex T-F domain, reports on WSJ0-2mix a strong SI-SDRi at 21.1 dB, which is comparable to the top results achievable by modern time-domain models.
It also shows stronger cross-corpus robustness than representative time-domain models.
Following DPRNN~\cite{Luo2020} and DPTNet~\cite{Chen2020DPTnet, Dang2022DPTFSNet}, it leverages a modern dual-path architecture but takes in a complex T-F spectrogram as input~\cite{Le2021DPCRN, Dang2022DPTFSNet}, and uses the transformer module proposed in DPTNet~\cite{Chen2020DPTnet} to model cross-frequency information in each frequency-scanning path and cross-frame information in each time-scanning path.
Although TFPSNet claims to operate in the T-F domain~\cite{Yang2022TFPSNet}, it closely follows the encoder-separator-decoder scheme~\cite{Luo2019} widely adopted in many TasNet variants: it (1) uses a one-dimensional (1D) convolution (Conv1D) layer followed by rectified linear units to encode the RI components of each T-F unit to a higher-dimensional embedding with non-negative values; (2) uses a dual-path separator to produce a non-negative mask to mask the embeddings for separation; and (3) applies a fully-connected layer at each T-F unit to predict the RI components of each speaker.
Its performance, however, does not surpass contemporary time-domain models~\cite{Qian2022, Rixen2022}, even with an advanced DNN architecture.

\begin{table}[t]
\scriptsize
\centering
\sisetup{table-format=2.2,round-mode=places,round-precision=2,table-number-alignment = center,detect-weight=true,detect-inline-weight=math}
\caption{Comparison with other systems on WSJ0-2mix.}
\vspace{-0.2cm}
\label{comparison_with_others}
\setlength{\tabcolsep}{3pt}
\begin{tabular}{
cc
S[table-format=4,round-precision=0]
S[table-format=3.1,round-precision=1]
S[table-format=2.1,round-precision=1]
S[table-format=2.1,round-precision=1]
}
\toprule

Systems & Domain & {Year} & {\#params (M)} & {SI-SDRi (dB)}  & {SDRi (dB)}\\

\midrule

DPCL++~\cite{Isik2016} & T-F & 2016 & 13.6 & 10.8 & {-} \\
uPIT-BLSTM-ST~\cite{Kolbak2017} & T-F & 2017 & 92.7 & {-} & 10.0 \\
ADANet~\cite{Chen2017a} & T-F & 2018 & 9.1 & 10.4 & 10.8 \\
WA-MISI-5~\cite{WZQe2eMISI2018} & T-F & 2018 & 32.9 & 12.6 & 13.1 \\
Sign Prediction Net~\cite{Wang2019Trigonometric} & T-F & 2019 & 56.6 & 15.3 & 15.6 \\
Conv-TasNet~\cite{Luo2019} & Time & 2019 & 5.1 & 15.3 & 15.6 \\
Deep CASA~\cite{Liu2019DeepCASA} & T-F & 2019 & 12.8 & 17.7 & 18.0 \\
Conv-TasNet-MBT~\cite{Lam2020MBT} & Time & 2020 & 8.8 & 15.6 & {-} \\
FurcaNeXt~\cite{Shi2019FurcaNeXt} & Time & 2020 & 51.4 & {-} & 18.4 \\
SUDO RM -RF~\cite{Tzinis2020} & Time & 2020 & 2.6 & 18.9 & {-} \\
DPRNN~\cite{Luo2020} & Time & 2020 & 2.6 & 18.8 & 19.0 \\
Gated DPRNN~\cite{Nachmani2020} & Time & 2020 & 7.5 & 20.1 & 20.4 \\
DPTNet~\cite{Chen2020DPTnet} & Time & 2020 & 2.7 & 20.2 & 20.6 \\
DPTCN-ATPP~\cite{Zhu2021} & Time & 2021 & 4.7 & 19.6 & 19.9 \\
SepFormer~\cite{Subakan2021} & Time & 2021 & 26.0 & 20.4 & 20.5 \\
Sandglasset~\cite{Lam2021} & Time & 2021 & 2.3 & 20.8 & 21.0 \\
Wavesplit~\cite{Zeghidour2020} & Time & 2021 & 29.0 & 21.0 & 21.2 \\
TFPSNet~\cite{Yang2022TFPSNet} & T-F & 2022 & 2.7 & 21.1 & 21.3 \\
MTDS (DPTNet)~\cite{Qian2022} & Time & 2022 & 4.0 & 21.5 & 21.7 \\
SFSRNet~\cite{Rixen2022} & Time & 2022 & 59.0 & 22.0 & 22.1 \\ 
QDPN~\cite{Rixen2022QDPN} & Time & 2022 & 200.0 & 22.1 & {-} \\ 

\midrule

TF-GridNet & T-F & 2022 & 14.4 & \bfseries 23.4 & \bfseries 23.5 \\ %

\bottomrule
\end{tabular}
\vspace{-0.5cm}
\end{table}

In this context, our study makes the following contributions to improve the performance of complex T-F domain approaches:
\begin{itemize}[leftmargin=*,noitemsep,topsep=0pt]
\item We propose to use complex spectral mapping for speaker separation in anechoic conditions.
Complex spectral mapping, initially proposed in~\cite{Williamson2016, Fu2017, Tan2020}, has shown strong potential on noisy-reverberant speech separation when combined with modern DNN architectures and loss functions~\cite{Wang2020CSMDereverbJournal, Wang2021LowDistortion, Wang2020css, Wang2021FCPjournal, Wang2021compensation, Tan2022NSF}, exhibiting strong robustness to noise and reverberation in both single- and multi-microphone conditions.
Its potential on anechoic speaker separation, however, has not been studied, especially in an era when time-domain models, which usually perform masking in learned embedding space, have become so popular and dominant on this task.
This paper is the first study to explore this direction.
\item We propose a novel DNN architecture, named TF-GridNet, for speech separation.
It operates in the complex T-F domain to model the spectro-temporal patterns in two-dimensional (2D), grid-like spectrograms.
Besides refining TFPSNet~\cite{Yang2022TFPSNet}, our study adds a full-band self-attention path for dual-path models to leverage cross-frame global information, resulting in a multi-path model.
\item Building upon the popular SI-SDR loss~\cite{Luo2019, LeRoux2018a}, we devise a novel time-domain loss term to encourage the summation of estimated sources to be close to the mixture.
\end{itemize}
Without using data augmentation and dynamic mixing, on WSJ0-2mix we obtain 23.4 dB SI-SDRi, which significantly surpasses the previous best (at 22.1 dB) attained by time-domain approaches~\cite{Rixen2022QDPN} and is slightly better than a theoretical SI-SDRi upper bound (at 23.1 dB) suggested in~\cite{sepit} for models with non-overlapping analysis filterbanks.
Both of these indicate the strong potential of complex T-F domain approaches also for anechoic speaker separation.
The code of TF-GridNet has been released in the ESPnet-SE++ toolkit~\cite{Lu2022ESPNetSE++}.

\vspace{-0.1cm}
\section{Proposed Algorithms}\label{proposedalgorithm}
\vspace{-0.1cm}

\begin{figure}
  \centering  
  \includegraphics[width=8.5cm]{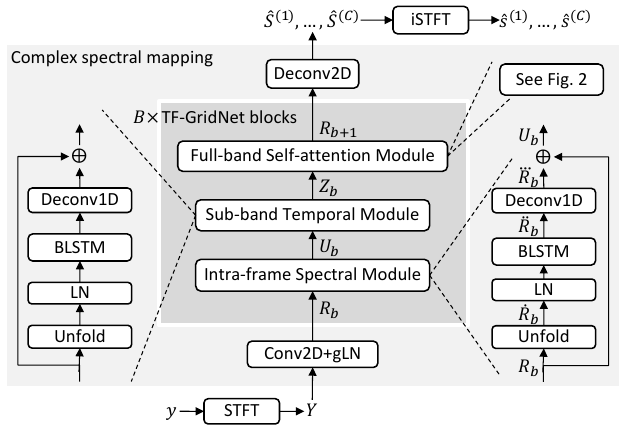}
  \vspace{-0.1cm}
  \caption{Overview of proposed system.}
  \label{system_overview}
  \vspace{-0.4cm}
\end{figure}

Given a $C$-speaker mixture recorded in anechoic conditions, the physical model in the time domain can be formulated as $y[n]=\sum_{c=1}^C s^{(c)}[n]$, where $y$ denotes the mixture and $s^{(c)}$ source $c$, and $n$ indexes $N$ time samples.
In the short-time Fourier transform (STFT) domain, the physical model is formulated as $Y(t,f)=\sum_{c=1}^C S^{(c)}(t,f)$, where $Y$ and $S^{(c)}$ respectively denote the complex spectra of $y$ and $s^{(c)}$, $t$ indexes $T$ frames, and $f$ indexes $F$ frequencies.
$C$ is assumed known in this study.
Our goal is to recover each source $s^{(c)}$ based on $y$.
An overview of the proposed system is provided in Fig.~\ref{system_overview}.
This section describes complex spectral mapping, DNN architectures, and loss functions.

\vspace{-0.2cm}
\subsection{Complex Spectral Mapping}

Our DNNs are trained to perform complex spectral mapping~\cite{Williamson2016, Fu2017, Tan2020, Wang2020CSMDereverbJournal, Wang2021LowDistortion, Wang2020css, Wang2021FCPjournal, Tan2022NSF}, where the RI components of $Y$ are concatenated as input features to predict the RI components of each speaker $S^{(c)}$.
The loss function, described later in Section~\ref{loss}, is defined based on the re-synthesized time-domain signals of the predicted RI components.
Our system is non-causal.
We normalize the sample variance of the time-domain mixture to one and scale each clean source using the same scaling factor during training. %

\vspace{-0.2cm}
\subsection{TF-GridNet}

Fig.~\ref{system_overview} shows the proposed system and Table~\ref{summary_hyperparam} summarizes the notations of the hyper-parameters in TF-GridNet.
Given an input tensor with shape $2\times T\times F$, where $2$ is because we stack RI components, we first use a 2D convolution (Conv2D) with a $3\times 3$ kernel followed by global layer normalization (gLN)~\cite{Luo2019} to compute a $D$-dimensional embedding for each T-F unit, obtaining a $D\times T\times F$ tensor.
Next, we feed the tensor to $B$ TF-GridNet blocks, each containing an intra-frame spectral module, a sub-band temporal module, and a full-band self-attention module, to gradually leverage local and global spectro-temporal information to refine the T-F embeddings.
At last, a 2D deconvolution (Deconv2D) with $2C$ output channels and a $3\times 3$ kernel followed by linear activation is used to obtain the predicted RI components, and inverse STFT (iSTFT) is applied for signal re-synethesis.
The rest of this section details the three modules in each TF-GridNet block.

\vspace{-0.25cm}
\subsubsection{Intra-frame Spectral Module}

In the intra-frame spectral module, we view the input tensor $R_b \in \RR^{D\times T \times F}$ to the $b^{\text{th}}$ block as $T$ separate sequences, each with length $F$, and use a one-layer bidirectional long short-term memory (BLSTM) to model the local spectral information within each frame.
We first use the \textit{torch.unfold} function~\cite{Paszke2019} with kernel size $I$ and stride size $J$ to stack nearby embeddings at each step, after we zero-pad the frequency dimension to $F'= \lceil \frac{F-I}{J} \rceil \times J + I$:
\begin{align}%
\dot{R}_b = \big[\text{Unfold}(R_b[:,t,:]), \,\text{for}\,\,t=1,\dots,T\big] \in \RR^{(I\times D)\times T \times (\frac{F'-I}{J}+1)}. \nonumber
\end{align}
Note that $J$ can be larger than one so that the sequence length and the amount of computation can be reduced.
We then apply layer normalization (LN) along the channel dimension (i.e., the first dimension) of $\dot{R}_b$, and a one-layer BLSTM with $H$ units in each direction is used to model the inter-frequency information within each frame:
\begin{align}%
\ddot{R}_b = \big[\text{BLSTM}\big(\text{LN}(\dot{R}_b)[:,t,:]\big), \,\text{for}\,\,t=1,\dots,T\big] \in \RR^{2H\times T \times (\frac{F'-I}{J}+1)}. \nonumber
\end{align}
After that, a 1D deconvolution (Deconv1D) layer with kernel size $I$, stride size $J$, input channel $2H$ and output channel $D$ (and without subsequent normalization and non-linearity) is applied to the hidden embeddings of the BLSTM:
\begin{align}%
\dddot{R}_b = \big[\text{Deconv1D}(\ddot{R}_b[:,t,:]), \,\text{for}\,\,t=1,\dots,T\big] \in \RR^{D\times T \times F'}. \nonumber
\end{align}
After removing the zero paddings, we add this tensor to the input tensor via a residual connection to produce the output tensor:
$U_b = \dddot{R}_b[:,:,:F] + R_b \in \RR^{D\times T \times F}$.

\vspace{-0.25cm}
\subsubsection{Sub-band Temporal Module}

In the sub-band temporal module, the procedure is almost the same as that in the intra-frame spectral module.
The only difference is that the input tensor $U_b \in \RR^{D\times T\times F}$ is viewed as $F$ separate sequences, each with length $T$, and a BLSTM is used to model temporal information within each sub-band.
The output tensor is denoted as $Z_b \in \RR^{D\times T\times F}$.

\vspace{-0.25cm}
\subsubsection{Full-band Self-attention Module}

\begin{figure}
  \centering  
  \includegraphics[width=8.5cm]{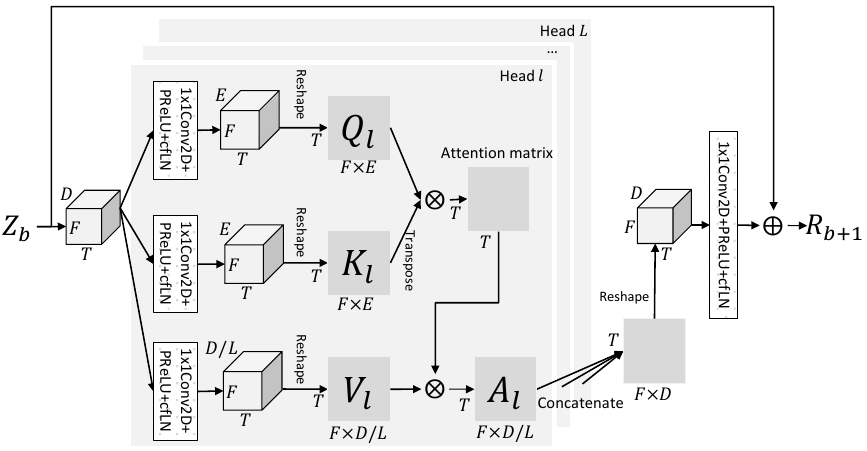}
  \vspace{-0.1cm}
  \caption{Proposed full-band self-attention module.}
  \label{self_attentenion_module}
  \vspace{-0.1cm}
\end{figure}

\begin{table}[t]
\scriptsize
\centering
\sisetup{table-format=2.2,round-mode=places,round-precision=2,table-number-alignment = center,detect-weight=true,detect-inline-weight=math}
\caption{Summary of model hyper-parameters.}
\vspace{-0.25cm}
\label{summary_hyperparam}
\begin{tabular}{
cc
}
\toprule

Symbols & Description  \\

\midrule

$D$ & Embedding dimension for each T-F unit  \\
$B$ & Number of TF-GridNet blocks  \\
\midrule
$I$ & Kernel size for Unfold and Deconv1D \\
$J$ & Stride size for Unfold and Deconv1D \\
$H$ & Number of hidden units in BLSTMs in each direction  \\
\midrule
$E$ & \begin{tabular}{@{}c@{}}Number of output channels in $1\times 1$ Conv2D to\\obtain query and key tensors in self-attention module\end{tabular} \\
$L$ & Number of heads in self-attention  \\

\bottomrule
\end{tabular}
\vspace{-0.5cm}
\end{table}

In the full-band self-attention module (illustrated in Fig.~\ref{self_attentenion_module}), given $Z_b$ produced by the sub-band temporal module, we first compute frame-level embeddings based on the T-F embeddings within each frame, and then use whole-sequence self-attention on these frame embeddings to capture long-range global information.
The motivation is that the intra-frame and sub-band BLSTMs can only model local information within each frame and each frequency, and a whole-sequence self-attention module enables each frame to attend to any frames of interest to exploit long-range information. 
Our self-attention module follows the attention mechanism proposed in~\cite{Liu2020Attn, Pandey2021}, which is used with U-Net for music source separation and speech denoising.
The key differences include (1) we use multi-head attention instead of single-head attention; and (2) we use the attention mechanism with dual-path models for speaker separation.

In detail, the self-attention module has $L$ heads, and, in each head $l$, we apply $1\times 1$ Conv2D followed by PReLU, LN along the channel and frequency dimensions (denoted as cfLN), and reshape layers to respectively obtain the 2D query $Q_l \in \RR^{T\times (F\times E)}$, key $K_l \in \RR^{T\times (F\times E)}$ and value $V_l \in \RR^{T\times (F\times D/L)}$ tensors.
The Conv2D layers for obtaining the query and key tensors both have $E$ output channels, leading to $F \times E$-dimensional query and key vectors at each frame after stacking the T-F embeddings within each frame, and similarly the Conv2D layer for computing the value tensor has $D/L$ output channels, resulting in an $F \times D/L$-dimensional value vector at each frame.
All the three $1\times 1$ Conv2D layers has $D$ input channels.
Following~\cite{Vaswani2017}, the attention output $A_l \in \RR^{T\times (F\times D/L)}$ is computed as
\begin{align}%
A_l = \text{softmax}(\frac{Q_l K_l^{\T}}{\sqrt{F\times E}})V_l. \nonumber
\end{align}
We then concatenate the attention outputs of all heads along the second dimension, reshape it back to $D\times T\times F$, apply $1\times 1$ Conv2D with $D$ input and $D$ output channels followed by PReLU and cfLN to aggregate cross-head information, and add it to the input tensor $Z_b$ via a residual connection to obtain the output tensor $R_{b+1}$, which is fed into the next TF-GridNet block.

This self-attention mechanism has two major advantages.
First, it only introduces a negligible number of parameters through the Conv2D layers.
Second, it operates at the frame level and the memory cost on attention matrices is $\mathcal{O}(B\times L \times T^2)$.
In contrast, TFPSNet~\cite{Yang2022TFPSNet} applies multi-head self-attention in each path-scanning module, and the memory cost on attention matrices is $\mathcal{O}\big(B\times L \times F \times T^2\big) + \mathcal{O}\big(B\times L \times T \times F^2\big)$, which is annoyingly high.

\vspace{-0.2cm}
\subsection{Loss Functions}\label{loss}

Our models are trained with utterance-level PIT~\cite{Kolbak2017}.
The loss function follows the SI-SDR loss~\cite{LeRoux2018a, Luo2019}, but with two differences.

First, in the paper proposing the SI-SDR metric~\cite{LeRoux2018a}, there are two equivalent formulations of SI-SDR, one scaling the \textit{target} to equalize its energy level with that of the estimate and the other instead scaling the \textit{estimate}.
The SI-SDR loss used in the seminal %
Conv-TasNet study~\cite{Luo2019} and almost all the follow-up studies employ the former formulation, while our study uses the latter, i.e.,
\begin{align}\label{sinrloss}
\mathcal{L}_{\text{SI-SDR-SE}} = - \sum_{c=1}^{C} 10\,\text{log}_{10} \frac{\| s^{(c)} \|_2^2}{\| \hat{\alpha}^{(c)}\hat{s}^{(c)} - s^{(c)} \|_2^2},
\end{align}
where $\hat{s}^{(c)}$ denotes the re-synthesized signal based on the predicted RI components for speaker $c$, $\hat{\alpha}^{(c)}={{\text{argmin}}}_{\alpha}\,\| \alpha \hat{s}^{(c)} - s^{(c)} \|_2^2=(\hat{s}^{(c)})^{\T}s^{(c)}/(\hat{s}^{(c)})^{\T}\hat{s}^{(c)}$, and the ``SE'' in $\mathcal{L}_{\text{SI-SDR-SE}}$ means ``scaling estimate''. We observe that this loss leads to faster convergence and very similar performance, compared with the former.

Second, the latter formulation motivates us to add a mixture-constraint (MC) loss between the mixture and the summation of the scaled estimated sources, defined, following~\cite{Wang2021FCPjournal}, as
\begin{align}\label{sinrloss+mc}
\mathcal{L}_{\text{SI-SDR-SE+MC}} = \mathcal{L}_{\text{SI-SDR-SE}} + \frac{1}{N} \Big\| \sum_{c=1}^{C} \hat{\alpha}^{(c)}\hat{s}^{(c)} - y \Big\|_1,
\end{align}
where the sample variance of $y$ has been normalized to one beforehand.
The second term results in better separation in our experiments.
It is motivated by a trigonometric perspective~\cite{Wang2019Trigonometric} in source separation, which suggested that constraining the separated sources to sum up to the mixture can lead to better phase estimation.
It should be noted that, at run time, $\sum_{c=1}^{C} \hat{\alpha}^{(c)}\hat{s}^{(c)}$ would not equal $y$.
This distinguishes our loss from mixture consistency~\cite{Wisdom2018MixtureConsistency}, which strictly enforces the separated sources to sum up to the mixture.
Our loss is also different from another mixture consistency loss proposed in~\cite{Zmolikova2021}, where the DNN is trained for real-valued magnitude masking without any phase estimation and the task is target speaker extraction based speech recognition in meeting scenarios. 

In Eq.~(\ref{sinrloss+mc}), we do not add a weight between the two terms for two reasons.
First, this can simplify the practical application of this loss function, as we can avoid a weight to tune.
Second, in modern speaker separation systems~\cite{WDLreview}, it is not uncommon for the SI-SDRi to surpass 10 dB, and when the sample variance of the input mixture has been normalized to one (which is the case in our study), the second term in our experiments has a very small scale (less than 0.01 when the models are converged).
This way, the second term would not dominate the overall loss, since it has a smaller influence compared with the first term, which is directly related to the final separation performance.

\vspace{-0.1cm}
\section{Experimental Setup}\label{setup}
\vspace{-0.1cm}

We validate the proposed algorithms on WSJ0-2mix~\cite{Hershey2016}, the most popular dataset to date to benchmark monaural talker-independent speaker separation algorithms in anechoic conditions.
It contains 20,000 ($\sim$30h), 5,000 ($\sim$10h), and 3,000 ($\sim$5h) two-speaker mixtures respectively in its training, validation and test sets.
The clean utterances are sampled from the WSJ0 corpus.
The speakers in the training and validation sets are not overlapped with the speakers in the test set.
The two utterances in each mixture are fully overlapped, and their relative energy level
is sampled uniformly from the range $[-5, 5]$\,dB.
The sampling rate is 8 kHz. %

The STFT window size is 32 ms and hop size 8 ms.
The square-root Hann window is used as the analysis window.
A 256-point discrete Fourier transform is applied to extract 129-dimensional complex spectra at each frame.
We use $B=6$ blocks and $E$ %
is set to $4$ for 8 kHz (see Table~\ref{summary_hyperparam} for their definitions).
This way, the dimension of frame-level embeddings (i.e., $F\times E$) used for self-attention is reasonable.
In each epoch, we sample a $4$-second segment from each mixture for model training.
Adam is used as the optimizer.
The norm for gradient clipping is set to $1$.
The learning rate starts from $0.001$ and is halved if the validation loss does not improve in 3 epochs.
SI-SDRi~\cite{LeRoux2018a} and SDRi~\cite{Vincent2006a} are used as the evaluation metrics, following previous studies.
The mixture SI-SDR is $0$\,dB and the mixture SDR $0.2$\,dB.
The number of model parameters is reported in million (M).

\section{Evaluation Results}\label{results}

Table~\ref{result_fair} compares the results of TF-GridNet with DPRNN~\cite{Luo2020} and TFPSNet~\cite{Yang2022TFPSNet}.
All the models are trained with the loss in~(\ref{sinrloss}).
Each model has almost the same number of parameters and uses almost the same amount of computation.
This is realized by using BLSTMs in all the models and unifying the embedding dimension (or the bottleneck dimension in the cases of DPRNN and TFPSNet, which perform masking in the embedded space) to $64$ and the hidden dimension of BLSTMs to $128$.
For DPRNN, the window size is set to $2$ samples, hop size to $1$ sample, chunk size to $250$ frames, and overlap between consecutive chunks to 50\%, following the best configuration reported in~\cite{Luo2020}.
For TF-GridNet, we remove the full-band self-attention module in each block and set both $I$ and $J$ to $1$ for a fair comparison.
From row 1, 2 and 5, we can see that TF-GridNet with complex spectral mapping obtains better results.
Table~\ref{result_fair} also provides the results of using TF-GridNet with masking in row 3 and 4.
In row 3, we perform masking in the learned embedded space, following~\cite{Luo2019,Luo2020,Yang2022TFPSNet}.
We employ the same encoder-separator-decoder modules used in~\cite{Yang2022TFPSNet}, but replace their path-scanning modules with our intra-frame spectral modules and sub-band temporal modules.
In row 4, we use TF-GridNet for complex ratio masking~\cite{Williamson2016, Liu2019DeepCASA}.
After obtaining the output tensor of the Deconv2D module (see Fig.~\ref{system_overview}), we truncate the values in the tensor to the range $[-5,5]$ to obtain a complex mask, and then multiply it with the mixture spectrogram for separation.
From row 3, 4 and 5, we observe that using complex spectral mapping is better.

\begin{table}[t]
\scriptsize
\centering
\sisetup{table-format=2.2,round-mode=places,round-precision=2,table-number-alignment = center,detect-weight=true,detect-inline-weight=math}
\caption{Comparison between masking and mapping based on WSJ0-2mix.}
\vspace{-0.2cm}
\label{result_fair}
\setlength{\tabcolsep}{1.5pt}
\begin{tabular}{
S[table-format=1,round-precision=0]
l
c
S[table-format=1.1,round-precision=1]
S[table-format=2.1,round-precision=1]
}
\toprule

{Row} & Systems & Masking or Mapping? & {\#params (M)} & {SI-SDRi (dB)} \\

\midrule

1 & DPRNN~\cite{Luo2020} & Masking learned embeddings & 2.6 & 18.8 \\
2 & TFPSNet (BLSTM)~\cite{Yang2022TFPSNet} & Masking learned embeddings & 2.6 & 19.7 \\

\midrule

3 & TF-GridNet & Masking learned embeddings & 2.8 & 20.7 \\ %
4 & TF-GridNet & Complex ratio masking & 2.6 & 20.8 \\
5 & TF-GridNet & Complex spectral mapping & 2.6 & \bfseries 21.2 \\

\bottomrule
\end{tabular}
\vspace{-0.25cm}
\end{table}

Table~\ref{result_hyperparam} presents the SI-SDRi results of our models on WSJ0-2mix using different model configurations.
From row 1-4, we observe that, when the kernel size is sufficiently large (i.e., $I=8$), using the Unfold and Deconv1D mechanism together with a smaller embedding dimension (i.e., $D=16$) does not degrade the performance, compared with the configuration that uses a larger embedding dimension (i.e., $D=128$) but does not stack nearby T-F embeddings (i.e., $I=1$).
One key benefit of using the former setup is that the memory consumption is noticeably lower.
From row 4 and 5, we notice that the MC loss produces slight improvement from 21.6 to 21.8 dB.
From row 5-7, we can see that enlarging the model size by increasing the embedding dimension $D$ and the number of hidden units $H$ in BLSTMs produces clear improvement.
The results in row 7, 8 and 9 indicate that the self-attention module is helpful (22.9 and 22.6 vs. 22.5 dB), and using four attention heads is better than using just one (22.9 vs. 22.6 dB).
Further enlarging the model size in row 10 pushes up the SI-SDRi to 23.4 dB.

\begin{table}[t]
\scriptsize
\centering
\sisetup{table-format=2.2,round-mode=places,round-precision=2,table-number-alignment = center,detect-weight=true,detect-inline-weight=math}
\caption{Ablation SI-SDRi (dB) results on WSJ0-2mix.}
\vspace{-0.2cm}
\label{result_hyperparam}
\setlength{\tabcolsep}{2pt}
\begin{tabular}{
S[table-format=2,round-precision=0]
c
c
S[table-format=1,round-precision=0]
S[table-format=3,round-precision=0]
S[table-format=1,round-precision=0]
S[table-format=1,round-precision=0]
S[table-format=3,round-precision=0]
S[table-format=2.1,round-precision=1]
c
S[table-format=2.1,round-precision=1]
}
\toprule

{Row} & Systems & Use attention? & {$L$} & {$D$} & {$I$} & {$J$} & {$H$} & {\#params (M)} & {Loss} & {SI-SDRi} \\

\midrule

1 & TF-GridNet & \xmark & {-} & 64 & 1 & 1 & 128 & 2.6 & (\ref{sinrloss}) & 21.2 \\
2 & TF-GridNet & \xmark & {-} & 16 & 4 & 1 & 128 & 2.6 & (\ref{sinrloss}) & 20.5 \\
3 & TF-GridNet & \xmark & {-} & 128 & 1 & 1 & 128 & 3.6 & (\ref{sinrloss}) & 21.6 \\
4 & TF-GridNet & \xmark & {-} & 16 & 8 & 1 & 128 & 3.6 & (\ref{sinrloss}) & 21.6 \\

\midrule

5 & TF-GridNet & \xmark & {-} & 16 & 8 & 1 & 128 & 3.6 & (\ref{sinrloss+mc}) & 21.8 \\
6 & TF-GridNet & \xmark & {-} & 16 & 8 & 1 & 192 & 6.5 & (\ref{sinrloss+mc}) & 21.9 \\
7 & TF-GridNet & \xmark & {-} & 24 & 8 & 1 & 192 & 8.0 & (\ref{sinrloss+mc}) & 22.5 \\

\midrule

8 & TF-GridNet & \cmark & 1 & 24 & 8 & 1 & 192 & 8.0 & (\ref{sinrloss+mc}) & 22.6 \\
9 & TF-GridNet & \cmark & 4 & 24 & 8 & 1 & 192 & 8.0 & (\ref{sinrloss+mc}) & 22.9 \\
10 & TF-GridNet & \cmark & 4 & 32 & 8 & 1 & 256 & 14.4 & (\ref{sinrloss+mc}) & \bfseries 23.351562 \\

\bottomrule
\end{tabular}
\vspace{-0.4cm}
\end{table}

Table~\ref{comparison_with_others} compares the performance of our system with others' based on WSJ0-2mix.
Our model has a modest number of parameters compared with previous best models such as SepFormer~\cite{Subakan2021}, SFSRNet~\cite{Rixen2022} and QPDN~\cite{Rixen2022QDPN}.
One notable observation is that, compared with time-domain models, T-F domain models, since 2019, have been largely under-explored and under-represented for speaker separation in anechoic conditions, and the research community has been focused on developing time-domain models.
The recent TFPSNet study~\cite{Yang2022TFPSNet} reports a strong performance at 21.1 dB SI-SDRi, but the performance falls within the range of scores (i.e., $[20.0,22.1]$ dB SI-SDRi) that can be commonly reached by modern time-domain models.
Our study, for the first time since 2019, unveils that complex T-F domain models, with a contemporary DNN architecture, can surpass the performance of modern time-domain models by a large margin.

\vspace{-0.1cm}
\section{Conclusions}\label{conclusion}
\vspace{-0.1cm}

We have proposed TF-GridNet, a novel model operating in the complex T-F domain for monaural speaker separation.
By combining it with complex spectral mapping and a time-domain loss with a mixture constraint, we obtain state-of-the-art 23.4 dB SI-SDRi on WSJ0-2mix without using dynamic mixing.
We believe that our study will generate broad impact, as it shows the strong performance of T-F domain models even for anechoic speaker separation.
There are many follow-up directions to investigate, such as leveraging more advanced DNN architectures, and extensions to noisy-reverberant speech separation and to multi-channel conditions.

In closing, we emphasize that, through years of efforts, the performance on WSJ0-2mix has been largely saturated~\cite{sepit}, and the separated signals do not sound too much different after the SI-SDRi surpassed, based on our informal listening tests, $22.0$ dB.
Nonetheless, the findings in this study suggest that, at a minimum, T-F domain methods modeling complex representations, which implicitly perform phase estimation by simultaneously predicting target RI components, are not sub-optimal compared with time-domain approaches for the task of anechoic speaker separation.
The performance gap between these two approaches in earlier studies could be mainly due to differences in DNN architectures, rather than because of the use of over-complete learned filterbanks with very short window length. %

\vspace{-0.1cm}
\section{Acknowledgments}\label{ack}
\vspace{-0.1cm}

We would like to thank Dr. Wangyou Zhang at SJTU for generously sharing his reproduced code of TFPSNet~\cite{Yang2022TFPSNet}.

\bibliographystyle{IEEEtran}
{\footnotesize
\bibliography{references.bib}
}

\end{document}